\documentclass[a4paper,notitlepage]{article} 

\usepackage{subfigure,epsfig}
\usepackage{cite}
\usepackage{url}
\usepackage{amsmath}
\usepackage{fancyvrb}
\RequirePackage{geometry}
\usepackage[affil-it]{authblk}
\usepackage{color}

\newcommand{\CLs}{\ensuremath{CL_s}}
\newcommand{\CLsb}{\ensuremath{CL_{s+b}}}
\newcommand{\CLb}{\ensuremath{CL_b}}
\newcommand{\mup}{\ensuremath{\mu_{\text{up}}}}
\newcommand{\snom}{\ensuremath{s^{\text{nom}}}}

\newcommand{\s}{\ensuremath{s}}
\newcommand{\back}{\ensuremath{b}}
\newcommand{\bi}{\ensuremath{b_i}}
\newcommand{\backinom}{\ensuremath{\bi^{\text{nom}}}}
\newcommand{\nobs}{\ensuremath{N^{\text{obs}}}}
\newcommand{\n}{\ensuremath{N}}
\newcommand{\fsyst}{\ensuremath{h^{\text{syst}}}}
\newcommand{\lhood}{\ensuremath{{\cal L}}}
\newcommand{\lhoodm}{\ensuremath{{\cal L}_\text{m}}}
\newcommand{\dd}{\ensuremath{{\rm d}}}
\newcommand{\pval}{\text{p-value}}

\author{E. Busato}
\affil{{\small\it LPC Clermont-Ferrand, CNRS/IN2P3, Universit\'e Blaise Pascal, France}}

\begin{document}

\title{Equivalence between hybrid \CLs~and bayesian methods for limit setting}
\maketitle

\begin{abstract}
The relation between hybrid \CLs~and bayesian methods used for limit setting is discussed. It is shown that the two methods are equivalent in the single channel case even when the background yield is not perfectly known. Only counting experiments are considered in this document.
\end{abstract}






\section{Introduction}

Two common methods used for setting upper limits on the number of signal events arising from some process of interest are the hybrid \CLs~and the bayesian methods. It is known that, in the single channel case without systematic uncertainties, both methods are equivalent when a uniform prior is used (see for example \cite{cowan}). It is shown in this document that this equivalence extends to the case where the background yield is affected by systematic uncertainties. 

The document is organized as follows. An overview of \CLs~and bayesian methods used for limit setting is given in Sec. \ref{sec:Overview}. The equivalence between \CLs~and bayesian methods in the case of perfectly known signal and backgrounds is discussed in Sec. \ref{sec:EquivHybridCLsBayesian}. This equivalence has been known for a long time but is demonstrated again here as it serves as an introduction to the more complex case where systematic uncertainties are included. The case with systematic uncertainties is discussed in Sec. \ref{sec:EquivHybridCLsBayesianWUncertainties}.



\section{Overview of \CLs~and bayesian methods}
\label{sec:Overview}

A brief overview of \CLs~and bayesian methods used for limit setting is given in this section. Both methods make use of the measurement likelihood which, in the single channel and counting experiment case, can be written as 
\begin{equation}
\label{eq:Lhood}
\lhood\left(\mu;\n\right)=\frac{\left(\mu\s+\back\right)^{\n}}{\n!}e^{-\left(\mu\s+\back\right)}
\end{equation}
where 
\begin{itemize}
\item $\mu$ is the signal strength
\item $\s$ is the signal yield
\item $\back$ is the background yield
\item $\n$ is the observed number of events
\end{itemize}

In the case where several processes contribute to the background yield, $\back$ can be expressed as
\[\back=\sum\limits_i \bi\]
where $\bi$ is the yield from process $i$.

In the rest of this document, the number of events actually observed in the data or the pseudo-data will be denoted as $\nobs$ and the confidence level $\alpha$. The upper limit on the signal strength will be denoted as $\mup$.

\subsection{\CLs~method}

The \CLs~method \cite{0954-3899-28-10-313} requires, as any frequentist method, the definition of a test statistic and the determination of the distribution of this test under background and signal plus background hypothesis. In the single channel and counting experiment case, the test statistic can be chosen to be, with no loss of generality, the observed number of events $\n$\footnote{The use of the classical likelihood ratio $\lhood(\mu)/\lhood(\mu=0)$ or the profile likelihood ratio $\lhood(\mu)/\lhood(\hat{\mu})$ (where $\hat{\mu}$ is the maximum likelihood estimator of $\mu$) as test statistics is equivalent to the use of the observed number of events (these various choices are related by a simple change of variable). The observed number of events is therefore used for simplicity, its distribution being known (Eq. \ref{eq:Lhood}).}. The distribution of $\n$ is given by Eq. \ref{eq:Lhood}. The $\CLs$ quantity is defined as
\[\CLs=\frac{\CLsb}{\CLb}\]
where
\[\CLsb=P\left(\n\leq\nobs;\mu\right)=\sum\limits_{\n=0}^{\nobs}\lhood\left(\mu;\n\right)\]
and
\[\CLb=P\left(\n\leq\nobs;\mu=0\right)=\sum\limits_{\n=0}^{\nobs}\lhood\left(\mu=0;\n\right)\]

The upper limit $\mup$ is obtained by solving 
\begin{equation}
\label{eq:CLsUpperLimitDef}
\CLs\left(\mup\right)=\alpha
\end{equation}

\subsection{Bayesian method}

The bayesian method makes use of the posterior distribution given by
\[p\left(\mu\right)=\frac{\lhood\left(\mu;\nobs\right)\pi(\mu)}{\displaystyle\int\limits_{0}^{\infty}\lhood\left(\mu;\nobs\right)\pi(\mu)\dd\mu}\]
where $\pi(\mu)$ is the prior on $\mu$. In what follows, only the uniform prior case will be considered, thus
\[p\left(\mu\right)=\frac{\lhood\left(\mu;\nobs\right)}{\displaystyle\int\limits_{0}^{\infty}\lhood\left(\mu;\nobs\right)\dd\mu}\]

The upper limit $\mup$ is obtained by solving
\begin{equation}
\label{eq:BayesianUpperLimitDef}
\displaystyle\int\limits_{0}^{\mup}p\left(\mu\right)\dd\mu=1-\alpha
\end{equation}

\section{Equivalence of \CLs~and bayesian methods without systematic uncertainties}
\label{sec:EquivHybridCLsBayesian}

The equivalence between Eq. \ref{eq:CLsUpperLimitDef} and Eq. \ref{eq:BayesianUpperLimitDef} in the case where the signal $\s$ and the background $\back$ are perfectly known can be established as follows. The bayesian definition of $\mup$ can be written as
\[1-\alpha=\frac{\displaystyle\int\limits_{0}^{\mup}\lhood\left(\mu;\nobs\right)\dd\mu}{\displaystyle\int\limits_{0}^{\infty}\lhood\left(\mu;\nobs\right)\dd\mu}=\frac{\displaystyle\int\limits_{0}^{\mup}\left(\mu\s+\back\right)^{\nobs}e^{-\left(\mu\s+\back\right)}\dd\mu}{\displaystyle\int\limits_{0}^{\infty}\left(\mu\s+\back\right)^{\nobs}e^{-\left(\mu\s+\back\right)}\dd\mu}\]

The numerator and denominator of this equation can be expressed using the incomplete gamma function $\Gamma\left(n+1;\nu\right)=\displaystyle\int\limits_{\nu}^{\infty}x^ne^{-x}\dd x$, yielding
\begin{equation}
\label{eq:BayesDefWithIncompleteGamma}
\alpha=\frac{\Gamma\left(\nobs+1;\mup s+b\right)}{\Gamma\left(\nobs+1;b\right)}
\end{equation}

From the equality
\[\sum\limits_{\n=0}^{\nobs}\frac{\nu^\n}{\n!}e^{-\nu}=\frac{\Gamma\left(\nobs+1;\nu\right)}{\Gamma\left(\nobs+1\right)}\]
it can be seen that the numerator and denominator of Eq. \ref{eq:BayesDefWithIncompleteGamma} are respectively $\Gamma\left(\nobs+1\right)\CLsb\left(\mup\right)$ and $\Gamma\left(\nobs+1\right)\CLb$. Eq. \ref{eq:BayesDefWithIncompleteGamma} is therefore equivalent to the $\CLs$ definition of the upper limit (Eq. \ref{eq:CLsUpperLimitDef}).

\section{Equivalence of hybrid \CLs~and bayesian methods with systematic uncertainties}
\label{sec:EquivHybridCLsBayesianWUncertainties}

Consider now the case where signal and background yields are affected by systematic uncertainties. These uncertainties are accounted for by introducing as many nuisance parameters as there are sources of uncertainties. Let $\eta_j$ be the nuisance parameter for systematic $j$ (from now on the index $j$ runs overs systematic sources). The yields are now functions of the $\eta_j$ :
\begin{itemize}
\item $\s=\snom\times\prod\limits_{j}\fsyst_{j}\left(\eta_j\right)$
\item $\back=\sum\limits_{i}\bi=\sum\limits_{i}\backinom\times\prod\limits_{j}\fsyst_{ij}\left(\eta_j\right)$
\end{itemize}

In the above expressions, $\snom$ and $\backinom$ are the nominal signal and background yields and $\fsyst$ are functions describing the variation of the yields with the nuisance parameters. The exact form of $\fsyst$ does not matter in what follows. The only assumption made is that the same functions are used in the \CLs~and the bayesian cases. The full likelihood is
\begin{equation}
\label{eq:fullLhood}
\lhood(\mu,\{\eta_j\};\n)=\frac{\left(\mu\s+\back\right)^{\n}}{\n!}e^{-\left(\mu\s+\back\right)}\prod\limits_{j}g\left(\eta_j\right)
\end{equation}
where $g(\eta_j)$ is the constraint term (prior) for nuisance parameter $\eta_j$. Eq. \ref{eq:fullLhood} corresponds to the case of independent systematic uncertainties. If systematic uncertainties are not independent, the joint probability density function should be used instead of $\prod\limits_{j}g\left(\eta_j\right)$. Note however that this does not change the calculation and the conclusion reached at the end remains valid.

The hybrid \CLs~and bayesian methods account for the effect of systematic uncertainties by integrating the likelihood over nuisance parameters (marginalization). The marginal likelihood is
\[\lhoodm\left(\mu;\n\right)=\displaystyle\int\lhood(\mu,\{\eta_j\};\n)~\prod\limits_j\dd\eta_j\]

Incorporating $\lhoodm$ in Eq. \ref{eq:CLsUpperLimitDef} leads to
\[\alpha=\frac{\CLsb\left(\mup\right)}{\CLb}=\frac{\sum\limits_{\n=0}^{\nobs}\lhoodm\left(\mup;\n\right)}{\sum\limits_{\n=0}^{\nobs}\lhoodm\left(\mu=0;\n\right)}\]

Thus,
\[\alpha=\frac{\displaystyle\int\sum\limits_{\n=0}^{\nobs}\frac{\left(\mup\s+\back\right)^{\n}}{\n!}e^{-\left(\mup\s+\back\right)}\prod\limits_{j}g\left(\eta_j\right)\dd\eta_j}{\displaystyle\int\sum\limits_{\n=0}^{\nobs}\frac{\back^{\n}}{\n!}e^{-\back}\prod\limits_{j}g\left(\eta_j\right)\dd\eta_j}\]
or
\begin{equation}
\label{eq:muUpFromCLsWithUncert}
\alpha=\frac{\displaystyle\int \CLsb\left(\mup,\{\eta_j\}\right)\prod\limits_{j}g\left(\eta_j\right)\dd\eta_j}{\displaystyle\int \CLb\left(\{\eta_j\}\right)\prod\limits_{j}g\left(\eta_j\right)\dd\eta_j}
\end{equation}
where $\CLsb\left(\mup,\{\eta_j\}\right)$ and $\CLb\left(\{\eta_j\}\right)$ are the \pval s for fixed values of the nuisance parameters.

The bayesian definition of $\mup$ with uniform prior on $\mu$ is, on the other hand,
\[1-\alpha=\frac{\displaystyle\int\limits_{0}^{\mup}\lhoodm\left(\mu;\nobs\right)\dd\mu}{\displaystyle\int\limits_{0}^{\infty}\lhoodm\left(\mu;\nobs\right)\dd\mu}\]

Thus,
\[1-\alpha=\frac{\displaystyle\int\left[\displaystyle\int\limits_{0}^{\mup}\left(\mu\s+\back\right)^{\nobs}e^{-\left(\mu\s+\back\right)}\dd\mu\right]\prod\limits_{j}g\left(\eta_j\right)\dd\eta_j}{\displaystyle\int\left[\displaystyle\int\limits_{0}^{\infty}\left(\mu\s+\back\right)^{\nobs}e^{-\left(\mu\s+\back\right)}\dd\mu\right]\prod\limits_{j}g\left(\eta_j\right)\dd\eta_j}\]

The terms between brackets in the numerator and denominator can be expressed using the incomplete gamma function as in Sec. \ref{sec:EquivHybridCLsBayesian}:
\[\begin{split}
1-\alpha&=\frac{\displaystyle\int\frac{\Gamma\left(\nobs+1;b\right)-\Gamma\left(\nobs+1;\mup s+b\right)}{s}\prod\limits_{j}g\left(\eta_j\right)\dd\eta_j}{\displaystyle\int\frac{\Gamma\left(\nobs+1;b\right)}{s}\prod\limits_{j}g\left(\eta_j\right)\dd\eta_j}\\
        &=\frac{\displaystyle\int\frac{\CLb\left(\{\eta_j\}\right)-\CLsb\left(\mup,\{\eta_j\}\right)}{s}\prod\limits_{j}g\left(\eta_j\right)\dd\eta_j}{\displaystyle\int\frac{\CLb\left(\{\eta_j\}\right)}{s}\prod\limits_{j}g\left(\eta_j\right)\dd\eta_j}
\end{split}
\]

The previous equation can be written as
\begin{equation}
\label{eq:muUpFromBayesWithUncert}
\alpha=\frac{\displaystyle\int\frac{\CLsb\left(\mup,\{\eta_j\}\right)}{s}\prod\limits_{j}g\left(\eta_j\right)\dd\eta_j}{\displaystyle\int\frac{\CLb\left(\{\eta_j\}\right)}{s}\prod\limits_{j}g\left(\eta_j\right)\dd\eta_j}
\end{equation}

Eq. \ref{eq:muUpFromCLsWithUncert} and Eq. \ref{eq:muUpFromBayesWithUncert} are equivalent if $\s$ is perfectly known ($\s=\snom$). This demonstrates the fact that the equivalence between hybrid \CLs~and bayesian methods extends to the case where the background yield has some systematic uncertainties associated to it.

\section{Conclusion}

Hybrid \CLs~and bayesian methods used for setting upper limits on signal strength have been discussed and compared. It has been shown that, for counting experiments with a single channel, both methods are equivalent if the signal yield is perfectly known.

\bibliographystyle{atlasBibStyleWoTitle}
\bibliography{Biblio}

\end{document}